\documentclass[preprint,12pt,nofootinbib]{revtex4-1}
\usepackage[paper=letterpaper,margin=1.5in]{geometry}

\usepackage{graphicx}\graphicspath{%
{graph/}}%
\usepackage{amssymb,amsmath,amsfonts}
\usepackage{time}
\usepackage{epstopdf}
\usepackage{hyperref}
\hypersetup{
  colorlinks,
  citecolor=magenta,
  linkcolor=blue}
\usepackage[T2A,T1]{fontenc} 
\usepackage[utf8]{inputenc} 
\usepackage{txfonts} 
\begin{document}

\begin{flushright}
{\normalsize
SLAC-PUB-16881\\
LCLS-II-TN-16-12\\
November 2016}
\end{flushright}

\vspace{.8cm}

\title{ Wakefields of a Beam near a Single Plate in a Flat Dechirper  \footnote[1]{Work supported by the U.S. Department of Energy, Office of Science, Office of Basic Energy Sciences, under Contract No. DE-AC02-76SF00515
} }

\author{Karl Bane\footnote[2]{kbane@slac.stanford.edu} and Gennady Stupakov}
\affiliation{SLAC National Accelerator Laboratory,Menlo Park, CA 94025}
\author{Igor Zagorodnov}
\affiliation{Deutsches Elektronen-Synchrotron, Notkestrasse 85, 22603 Hamburg, Germany}

\begin{center}
\end{center}



\maketitle

\section*{Introduction}


At linac-based, X-ray free electron lasers (FELs), 
there is interest in streaking the beam by inducing the transverse wakes in a flat dechirper, by passing the beam near to one of its two jaws~(see {\it e.g.} \cite{Novokhatski15}). For LCLS-II---as has already been done for LCLS-I---this way of using the dechirper will {\it e.g.} facilitate two-color and fresh slice schemes of running the FEL~\cite{Lutman16}. With the beam a distance from the near wall of say $b\sim0.25$~mm and from the far wall by $\gtrsim5$~mm, the second wall will no longer affect the results. The physics will be quite different from the two plate case: with two plates the impedance has a resonance spike whose frequency depends on the plate separation $2a$; in the single plate case this parameter no longer exists. 
Formulas for the longitudinal, dipole, and quadrupole wakes for a beam off-axis between two dechirper plates, valid for the range of bunch lengths of interest in an X-ray FEL, are given in Ref.~\cite{analytical}. By taking the proper limit, we can obtain the corresponding wakes for a beam close to one dechirper plate and far from the other. This is the task we perform in this note. 

\begin{figure}[htb]
\centering
\includegraphics[width=0.5\textwidth, trim=0mm 0mm 0mm 0mm, clip]{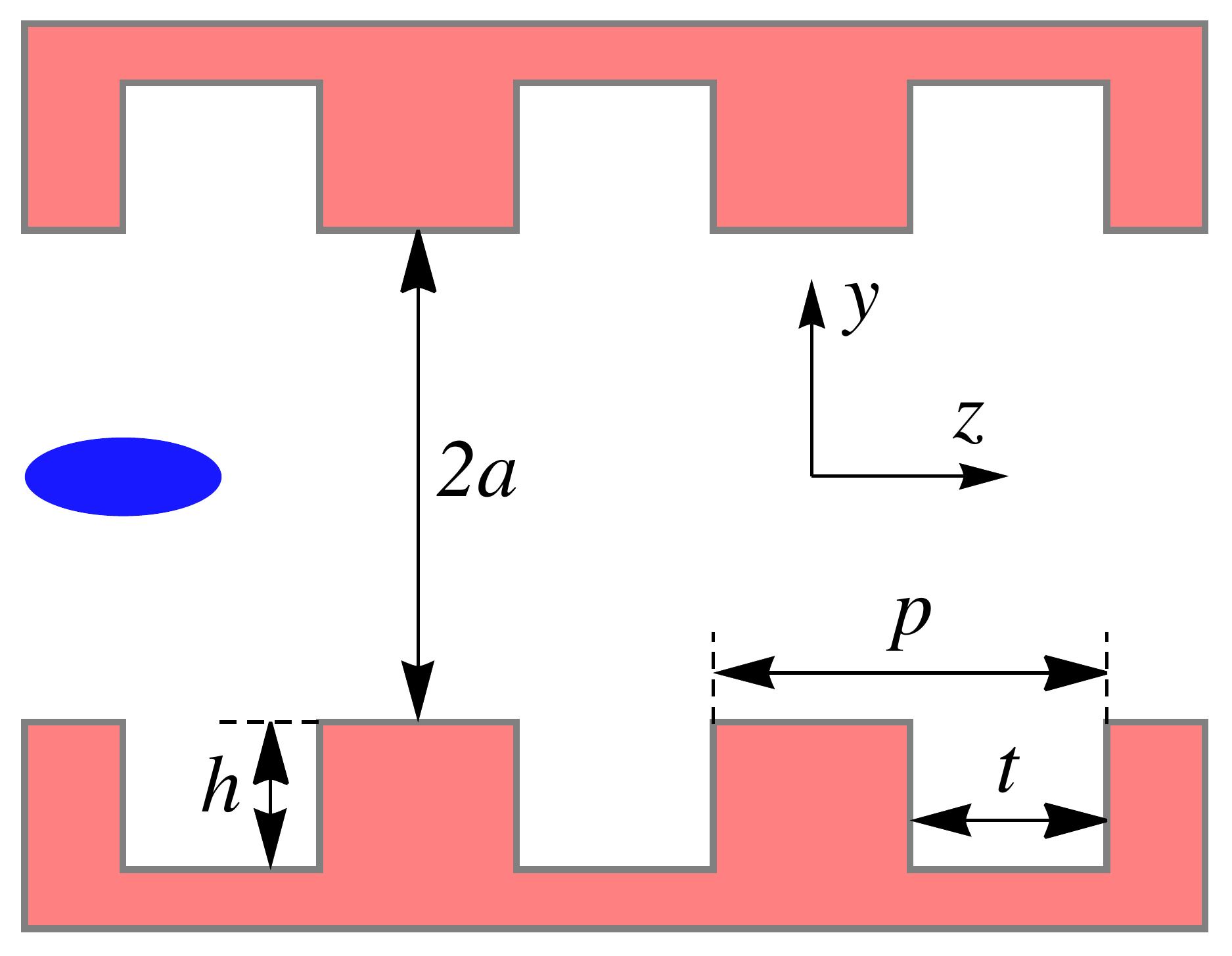}
\caption{Three corrugations of a vertical dechirper. A rectangular coordinate system is centered on the symmetry axis of the chamber. The $x$ axis points into the page. The blue ellipse represents an electron beam propagating along the $z$ axis.}
\label{fig:1}
\end{figure}

Fig.~\ref{fig:1} gives a sketch of a beam on axis (the blue ellipse) in a vertical dechiper consisting of two plates, showing the parameters corrugation depth $h$, longitudinal gap $t$, and period $p$; the separation of the plates is denoted by $2a$. If we move the lower jaw downwards by a large amount while leaving the location of the beam and the upper jaw fixed, then the beam will interact only with the upper jaw, which is the condition of interest here. Table~I gives the parameters of the RadiaBeam/SLAC dechirper that has been installed in LCLS-I, and that will be used in example calculations. Note that to serve as a good dechirper, it is required that $p\lesssim a$ and $h\gtrsim p$ (and, of course, $a\ll w$, with $w$ the plate width).

\begin{table}[hbt]
   \centering
   \caption{Selected  properties of the RadiaBeam/SLAC dechirper. The width is the $x$ extent of the plates.}
   \begin{tabular}{||l|c|c||}\hline 
        {Parameter name} & {Value}  &  Unit\\ \hline\hline 
       Period, $p$ & 0.5 & mm \\
       Longitudinal gap, $t$ & 0.25 & mm\\
       Full depth, $h$ & 0.5 & mm\\    
       Nominal half aperture, $a$       & 0.7  & mm \\   
       Plate width, $w$       &12  & mm \\      
     Plate length, $L$       &2  &m \\
         \hline \hline 
   \end{tabular}
   \label{table1_tab}
\end{table}

The wakefields in this note are given in Gaussian units. To convert to the MKS system, one needs to multiply by $(Z_0c/4\pi)$, with $Z_0=377$~$\Omega$ and $c$ is the speed of light. 

\section*{Wakefields}

\begin{figure}[htb]
\centering
\includegraphics[width=0.8\textwidth, trim=0mm 0mm 0mm 0mm, clip]{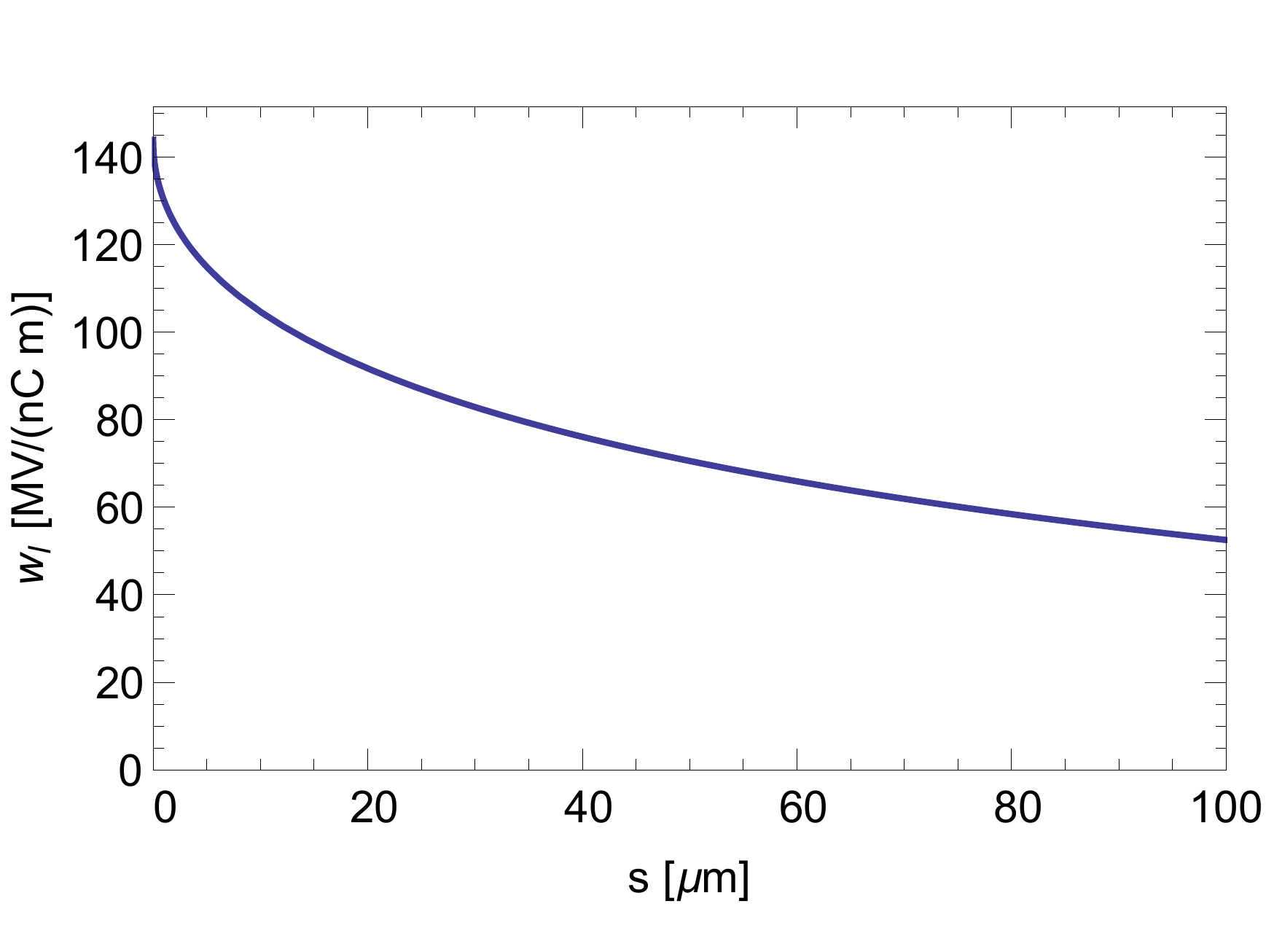}
\caption{Longitudinal wake of the dechirper with the beam $b=0.25$~mm from one plate, with the other plate far away.}
\label{wl_sp_fi}
\end{figure}

We start with the longitudinal wake for the case driving and test particles offset by $y$ from the axis of a dechirper with full gap $2a$ (Eqs.~14, 15, of~\cite{analytical}):
\begin{equation}
w_l(s,y)=\frac{\pi ^2}{ 4a^2}\sec^2\left(\frac{\pi y}{2a}\right)e^{-\sqrt{s/s'_{0l}}}\ ,\label{Wl_eq}
\end{equation}
with
\begin{equation}
s'_{0l}=\left(\frac{2a^2t}{\pi \alpha^2p^2}\right)\left[1+\frac{1}{3}\cos^2(\frac{\pi y}{2a})+(\frac{\pi y}{2a})\tan(\frac{\pi y}{2a})\right]^{-2}\ \label{s0l_eq}
\end{equation}
and $\alpha=1-0.465\sqrt{t/p}-0.070(t/p)$. 
To obtain the wake for the case of the two particles offset from one plate by distance $b$ (with the other plate far away), we let $y=a-b$ and then let  $a\rightarrow\infty$. We find that
\begin{equation}
w_l(s)=\frac{1}{b^2}e^{-\sqrt{s/s_{0l}}}
\end{equation}
with $s_{0l}=2b^2t/(\pi\alpha^2p^2)$. Here the corrugation parameters are period $p=0.5$~mm, longitudinal gap $t=0.25$~mm, and $\alpha=0.636$. With distance from wall $b=0.25$~mm, $s_{0l}=98$~$\mu$m. Fig.~\ref{wl_sp_fi} shows the longitudinal, point charge wake.

Note that the wake at the origin for one plate is $w_l(0^+)=1/b^2$, while for two plates, separated by $2a$, it was $w_l(0^+)=\pi^2/(4a^2)$.

\begin{figure}[htb]
\centering
\includegraphics[width=0.8\textwidth, trim=0mm 0mm 0mm 0mm, clip]{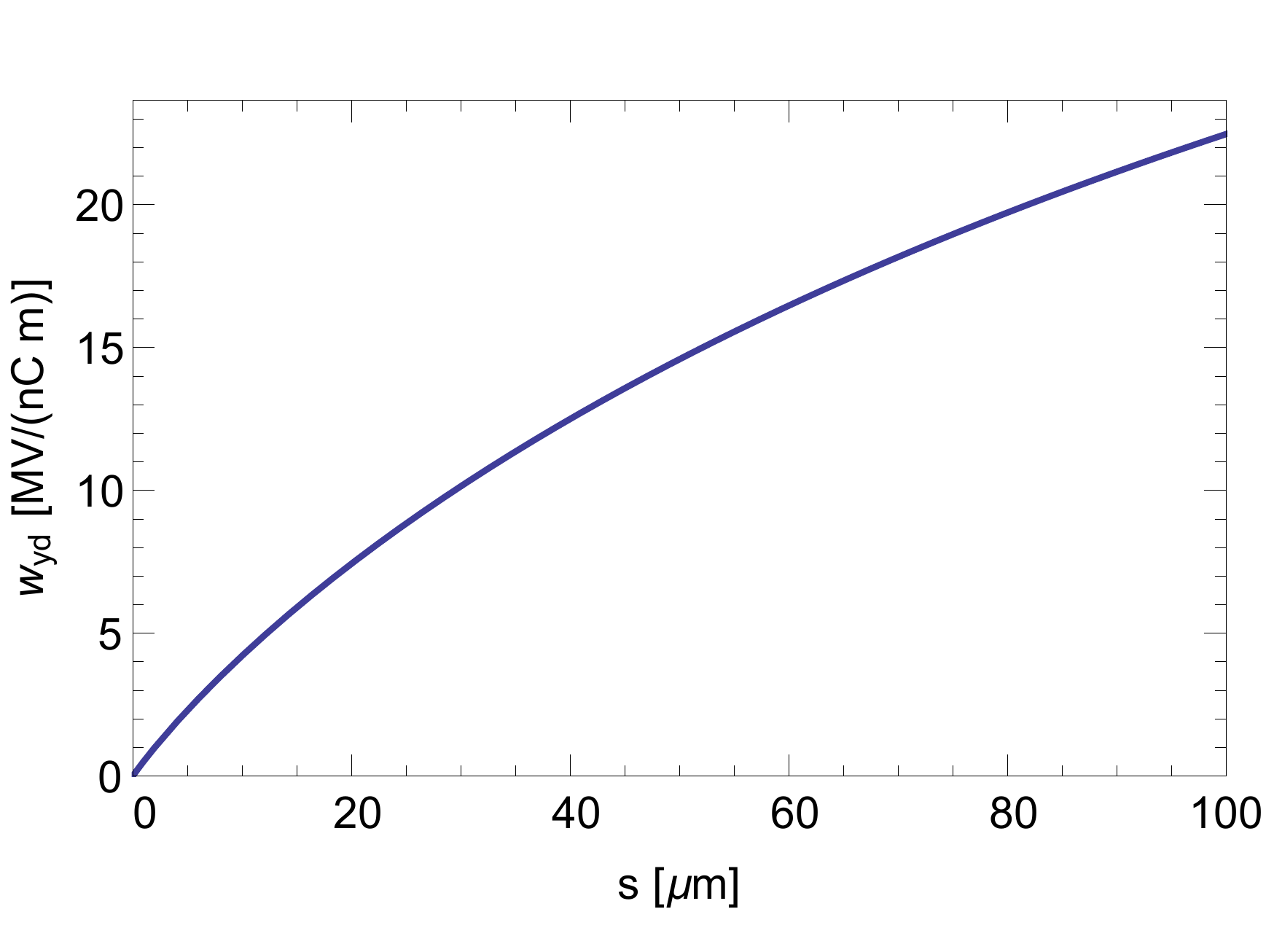}
\caption{Dipole wake of the dechirper with the beam $b=0.25$~mm from one plate, with the other plate far away.}
\label{wyd_sp_fi}
\end{figure}

We assume that the dechirper has its jaws parallel to the $x$-axis. With driving particle at $(x=0,y)$ and test particle at $(x,y+\Delta y)$, the transverse wakes of a flat dechirper are given by
\begin{eqnarray}
w_y(y)&=&w_{yd}(y)+\Delta y\, w_{yq}(y)\nonumber\\
w_x(x)&=&-xw_{yq}(x)\ ,
\end{eqnarray}
where $w_{yd}$ is the dipole and $w_{yq}$ the quadrupole wake component in the vicinity of $y$. For the dipole wake component we take Eqs.~33, 34, of Ref.~\cite{analytical}, again let $y=a-b$ and then let $a\rightarrow\infty$ we find that
\begin{equation}
w_{yd}(s)=\frac{2}{b^3}s_{0yd}\left[1-\left(1+\sqrt{\frac{s}{s_{0yd}}}\right)e^{-\sqrt{s/s_{0yd}}}\right]
\end{equation}
with $s_{0yd}=8b^2t/(9\pi\alpha^2p^2)$. Here  $s_{0yd}=44$~$\mu$m. Fig.~\ref{wyd_sp_fi} shows the dipole, point charge wake.

\begin{figure}[htb]
\centering
\includegraphics[width=0.8\textwidth, trim=0mm 0mm 0mm 0mm, clip]{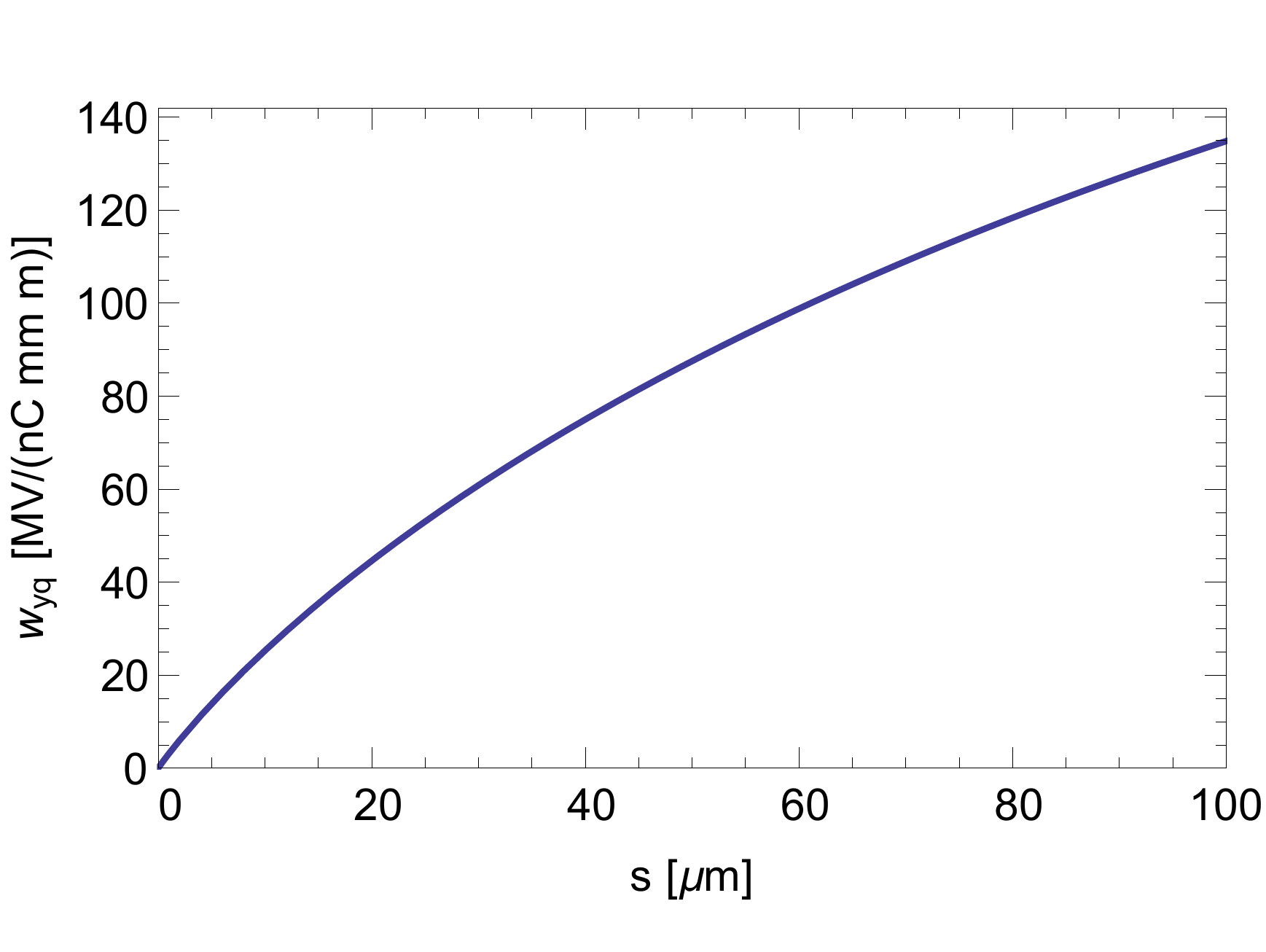}
\caption{Quad wake of the dechirper with the beam $b=0.25$~mm from one plate, with the other plate far away.}
\label{wyq_sp_fi}
\end{figure}

 For the quad wake component we follow the same procedure using Eqs.~36, 37, of Ref.~\cite{analytical} to obtain
\begin{equation}
w_{yq}(s)=\frac{3}{b^4}s_{0yq}\left[1-\left(1+\sqrt{\frac{s}{s_{0yq}}}\right)e^{-\sqrt{s/s_{0yq}}}\right]
\end{equation}
with $s_{0yq}=8b^2t/(9\pi\alpha^2p^2)$. Here  $s_{0yq}=44$~$\mu$m. Fig.~\ref{wyq_sp_fi} shows the quadrupole, point charge wake.

For an example LCLS calculation, consider the case of bunch offset $b=0.25$~mm; of a uniform beam distribution with charge $Q=300$~pC and peak current $I=1.5$~kA. The wake voltage induced in the tail of the bunch (of length $\ell=60$~$\mu$m) when passing one dechirper jaw is $V(\ell)=QLW_\lambda(\ell)$, where bunch wake is $W_\lambda(s)=-\ell^{-1}\int_0^s w_l(s')\,ds'$, and the length of the dechirper jaw is $L$. The induced voltage at the bunch tail is $V(\ell)=-52$~MV.
For the dipole wake, performing the same calculation we find that the kick in the tail of the bunch is $V_{yd}(\ell)=5.7$~MV. Finally, with the quad wake we have a defocussing that is largest in the bunch tail. The inverse focal length is given as $f_q^{-1}(s)=\pm eQLW_{\lambda q}(s)/E$, with $E$ the beam energy. With $E=6$~GeV, we obtain a focusing/defocussing in the bunch tail of $f_q^{-1}(\ell)=\pm5.7$~m$^{-1}$.

\end{document}